\documentstyle[prb,aps,multicol,epsfig]{revtex}
\tighten
\newcommand{\beq}{\begin{equation}}
\newcommand{\eeq}{\end{equation}}
\newcommand{\bea}{\begin{eqnarray}}
\newcommand{\eea}{\end{eqnarray}}

\begin{document}

\draft

\title{Complexation of DNA with positive
spheres: phase diagram of charge inversion and reentrant condensation}

\author{Toan T. Nguyen \and Boris I. Shklovskii}

\address{Theoretical Physics Institute, University of Minnesota, 116
  Church St. Southeast, Minneapolis, Minnesota 55455}

\maketitle

\vspace{.2cm}
\centerline{\today}

\begin{abstract}
The phase diagram of a water solution of DNA and oppositely charged
spherical macroions is studied. DNA winds around spheres 
to form beads-on-a-string complexes resembling the chromatin 10 nm fiber. 
At small enough concentration of spheres these "artificial chromatin" 
complexes are negative, while at large enough concentrations of spheres
the charge of DNA is inverted by the adsorbed spheres.
Charges of complexes stabilize their solutions. 
In the plane of concentrations of DNA and spheres the 
phases with positive and negative complexes
are separated by another phase, which contains the
condensate of neutral DNA-spheres complexes. Thus when the
concentration of spheres grows, DNA-spheres complexes
experience condensation and resolubilization (or reentrant condensation).
Phenomenological theory of the phase diagram of reentrant 
condensation and charge inversion is suggested.
Parameters of this theory are calculated 
by microscopic theory.
It is shown that an important part of the effect of
a monovalent salt on the phase diagram
can be described by the nontrivial renormalization of 
the effective linear charge density of DNA 
wound around a sphere, due to the Onsager-Manning condensation. 
We argue that our phenomenological phase diagram or
reentrant condensation
is generic to a large class of strongly asymmetric electrolytes.
Possible implication of these results for the natural
chromatin are discussed.

\end{abstract}

\begin{multicols} {2}

\section{Introduction}

\begin{figure}[htbp]
    \epsfxsize=9cm \centerline{\epsfbox{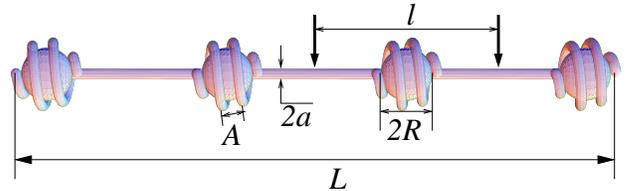}}
\caption{A beads-on-a-string structure of the complex of a long DNA
	 helix  with positive spheres. We assume that the number of turns, 
    $m$, of DNA necessary to neutralize a sphere is large, $m > 1$.}
\label{fig:bead}
\end{figure}

In the chromatin a long negative DNA
double helix winds around a positive
histone octamer to form a complex known
as the nucleosome\cite{Alberts}. Presumably due to
their Coulomb repulsion nucleosomes position
themselves equidistantly along the DNA helix, 
forming a periodic necklace or
a beads-on-a-string structure,
which is also called a 10 nm fiber.
The fact that the self assembly
of this structure is very sensitive to the salt concentration shows
that electrostatic forces are important.  In proper conditions, the 10
nm fiber self assembles (condenses)
into the 30 nm chromatin fiber, which is the
major building material of a chromosome\cite{Alberts}. 
It is interesting to
understand whether electrostatic forces alone are sufficient to form
the 10 nm fiber and determine the range of its stability.
This is one of the motivations to study theoretically and experimentally
a model of ``artificial chromatin", where
histone octamers (which themselves are in a complex equilibrium with
a population of individual histones) are replaced by identical
positive hard spheres (see Fig.  \ref{fig:bead}). 
In experimental realizations of ``artificial chromatin", spheres 
can be colloidal particles\cite{Sivan,Grant,Gotting},
micelles\cite{Dubin,Wang} or
dendrimers\cite{Safinya}. Other motivations are related to 
many applications of such complexes.
One of them is the gene therapy. Inversion of charge of DNA by its
complexation with positive spheres should facilitate its contact with a
usually
negative cell membrane making penetration into the cell more
likely. Another application we want to mention here is the possibility
of
making DNA guided nanowires\cite{Sivan} by complexation of metallic
spherical colloids with DNA. When DNA is replaced by synthetic 
polyelectrolyte (PE), other industrial applications emerge from 
pharmaceutics to protein extraction and purification.
In the previous paper\cite{Shklov003} we analytically considered
a worm-like negative PE (DNA) with the total
charge $-Q$ and uniformly charged positive spheres with charge
$q$. (Below we use the word DNA instead
of PE postponing generalizations to the end of the paper).
We studied their complexation in a water solution containing a
concentration $p$ of DNA molecules and a concentration $s$ of spheres
together with their monovalent counterions and a finite concentration
of monovalent salt which provides Debye-H\"{u}ckel screening with the
screening radius $r_s$. In the
Ref.~\onlinecite{Shklov003} we dealt only with large enough
concentrations $s$, neglecting entropy of free spheres.
We have shown that, at large $s$, the state of the system is
determined by the ratio $s/s_i(p)$, where $s_i(p)=q/pQ$ is
 isoelectric concentration
at which the total charge of DNA is equal to the total charge of spheres.
At $s < s_i(p)$ DNA winding around a
sphere overscreens it so that the net charge of the sphere becomes
negative. Negative spheres repel each other and form a periodic necklace.
In the opposite case $s > s_i(p)$,
more spheres bind to DNA than necessary to neutralize it,
thus inverting the charge of DNA.
The origin of this counterintuitive phenomenon of charge inversion is
the correlation induced attraction of a new sphere to a neutral complex:
a new sphere approaching a neutral complex pushes away
neighboring spheres, unwinds some of
DNA from them and winds it on itself.  In
other words, a new sphere creates an oppositely charged image in the
complex and binds to it. It is this additional attraction which
leads to charge inversion. In this case, the net charge of
each sphere is positive. Positive spheres repel each other and
also form a periodic necklace.

\begin{figure}
\epsfxsize=7.0cm \centerline{\epsfbox{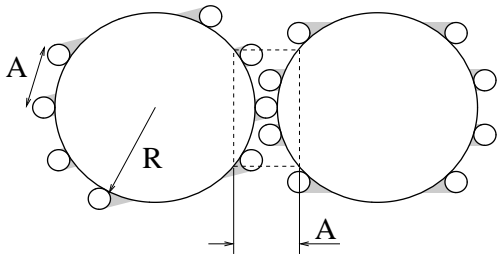}}
\caption{Cross section through the centers of two touching spheres
with worm-like (gray) PE wound around them.
At the place where two spheres touch each other
(the region bounded by the broken line)
the density of PE doubles which in turn
leads to a gain in the correlation energy of PE segments
wound around the spheres.}
\label{fig:2spheres}
\end{figure}

\begin{figure}
\epsfxsize=5.0cm \centerline{\epsfbox{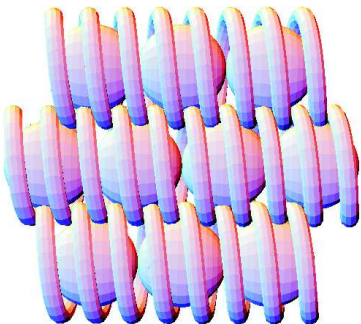}}
\caption{Almost neutral complexes condense into macroscopic bundles.}
\label{fig:bundle}
\end{figure}

Far from the isoelectric point, $s = s_i(p)$,
complexes have a beads-on-a-string
structure (see Fig.  \ref{fig:bead}).  They are strongly charged and
repel each other in the solution, so that solution is stable.
At the isoelectric point, $s = s_i(p)$, the
sphere net charge and the charge of the whole complex
simultaneously cross zero.
In the narrow vicinity of the
isoelectric point, the charges of complexes (and their Coulomb
repulsion) are so small that their
short range attraction between complexes
due to correlations of solenoids of DNA on
different spheres (see Fig.  \ref{fig:2spheres}) is able
to condense them. Complexes form large and weakly charged bundles (see Fig.
\ref{fig:bundle})
At a given concentration $p$ of DNA and growing concentration $s$ of
spheres, DNA molecules experience condensation (aggregation)
at some critical
concentration $s=s_c(p)$ below the isoelectric point $s=s_i(p)$, 
and remain in
condensed state until another concentration $s=s_d(p)$ above the
isoelectric
point, where the inverted positive charge of the complex
becomes so large that condensate
dissolves. The electrophoretic mobility changes sign at the isoelectric
point $s=s_i(p)$ located between $s_c(p)$ and $s_d(p)$.
Inversion of sign of electrophoretic
mobility of complexes of DNA with dendrimers was
recently reported in Ref. \onlinecite{Safinya}.

\begin{figure}
    \epsfxsize=7cm \centerline{\epsfbox{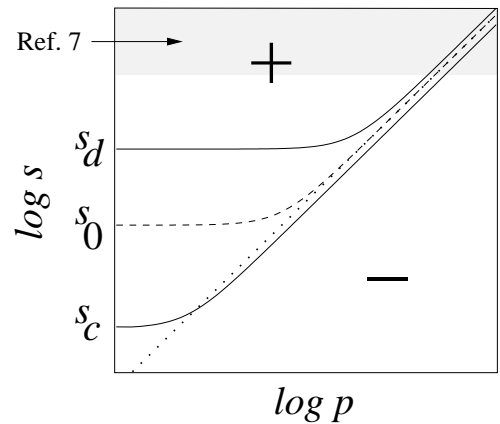}}
    \caption{Phase diagram of a water solution of long DNA
	molecules and 
	  positive spheres
      in the ($p$,$s$) plane. The dotted line corresponds
      to the isoelectric composition $s=s_i(p)$. 
	  The dashed line corresponds to
      the concentration of spheres $s = s_{0}(p)$,
      where an isolated DNA-spheres complex is
      neutral. The solid lines $s_c(p)$ and $s_d(p)$ 
	  define the external boundary of the region of
      existence of macroscopic aggregates.
      Plus and minus are the signs of the charge of free DNA-sphere
	  complexes and their aggregates
      above and below the dashed curve. The area of large $s$
      studied in previous paper\cite{Shklov003} is shown by gray.}
    \label{fig:phase}
\end{figure}

In Ref. \onlinecite{Shklov003}, we dealt only with large $s$
and concentrated mainly on the structure and charge of
free necklaces far from the condensation domain.
In this paper, we calculate
the dependence of the critical concentrations
$s_{c}(p)$ and $s_{d}(p)$ for all $(p,s)$ plane
and study details of the condensation domain.
Our main result is shown by the phase diagram
in Fig. \ref{fig:phase}. In this diagram, an aggregate of
DNA-spheres complexes exists only in the region surrounded by the lines
$s_c(p)$ and $s_d(p)$ (the two solid lines).
The domain of large enough concentrations $s$, studied
Ref.~\onlinecite{Shklov003} is shown by gray. 
In the gray area condensation domain is narrow and
encloses isoelectric line $s_{i}(p)$.
We call this part of phase diagram the ``neck".
At smaller $s$ and $p$, the condensation region is found to be
much wider so that we call it the ``body". The dash line
corresponds to the values of $s$ and $p$ at which complexes are
neutral. Notice that,
in agreement with result of Ref. \onlinecite{Shklov003},
this line is essentially the isoelectric line
for the ``neck", but it is high above the isoelectric line
for the ``body", where substantial fraction 
of spheres is free.

\begin{figure}
    \epsfxsize=9cm \centerline{\epsfbox{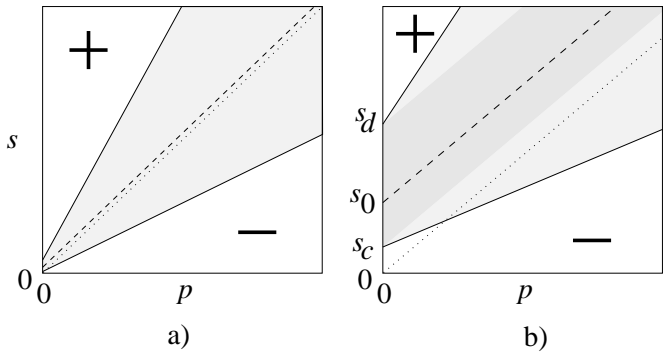}}
	\caption{a) Phase diagram of the mixture of long DNA helix and 
	spheres in linear scale. b) Domain of small $s$ and $p$ 
	is enlarged. On both plots, 
	the dotted and dashed lines and plus
	and minus signs have the same meaning as in
	Fig. \ref{fig:phase}.
	The areas of coexistence of aggregates and
	free complexes are shown by light gray.
	In the dark gray region, all DNA
	helices condense. It is shown only in plot b) because
	it should be very narrow in plot a). One can see that
	the transitions to and from the complete
	condensation are
	abrupt at small $p$ but become less steep with growing $p$.}
	\label{fig:lphase}
\end{figure}

The same diagram is shown in linear scale on Fig. \ref{fig:lphase}.
At large $s$ and $p$, the phase
diagram is extremely simple and centered around
the isoelectric line. 
When we zoom at smaller $s$ and $p$, symmetry with respect to
the isoelectric line disappears 
and we see the fine structure of the
condensation domain.  There are two light
gray regions
next to lines  $s_c(p)$ and $s_d(p)$, where condensate coexists with
free necklaces. Between them there is the dark gray region 
were practically all DNA and spheres are consumed by the
condensate and concentration of free necklaces is exponentially small.
We also study the change in the fraction of free
necklaces and their charge when we cross all these regions,
for example, by increasing $s$, while keeping
$p$ constant (see Fig. \ref{fig:fraction} below).
One of interesting results is that the inversion
of the necklace charge happens in a relatively narrow range of
$s$. 
In Figs. \ref{fig:phase} and \ref{fig:lphase}
we did not try to show that the lines $s_c(p)$ and
$s_d(p)$ should merge at extremely small $p$, where translational
entropy of DNA molecules
becomes important and as a result DNA complexes dissolve.
These lines may also merge again when $s$ and $p$ are so large that
the fraction of water in the mixture is significantly reduced.
Then our phase diagram can also be redrawn as a close loop
ternary miscibility diagram.

Bundles of complexes formed in the gray
area of phase diagram of Fig. 5 are almost
neutral in the sense that their charge is proportional not 
to the number of
aggregated complexes or volume of the bundle, but to the linear size or
surface area of bundle. The sign of the charge is however same as for
free complexes. It flips on the neutrality line
where charge inversion takes place for free complexes.

To derive these phase diagrams, we use two phenomenological
parameters. The first one, $\varepsilon$, is
the binding energy per sphere of the
DNA-spheres complex (necklace) in the
aggregate. The second one, $s_0$, is the concentration of spheres,
which is in equilibrium with neutral
DNA-spheres complexes. This concentration
separates domains of positive and negative complexes
at $p=0$.
These two phenomenological parameters can be extracted from the
experimental phase diagram. They can also be calculated from
microscopic theory.
In this paper we present comprehensive microscopic theory
of these parameters and study the
influence of salt
on their value. We show below that, generally speaking,
screening by salt increases the width of the neck and,
at strong screening, increases $s_0$ so that the
``body" grows at the expense of the ``neck".

It is known that the Onsager-Manning condensation
of monovalent counterions~\cite{Manning}
renormalizes linear charge density of free DNA helix from the 
bare value $-\eta$
to  critical Onsager-Manning density $-\eta_c = -Dk_BT/e$.
Thus one can ask which of the two one should
use to calculate charge of DNA molecule, $Q$, and the isoelectric
point $s_i$. We show that, because 
of the positive charge of the spheres surface, some DNA counterions
are released into solution. As a result,
the absolute value of the 
effective net linear density, $\eta^*$, of DNA
(which almost neutralizes a sphere)
decreases with decreasing
screening length $r_s$ from the bare density $\eta$ to the
Onsager-Manning critical density $\eta_c$ according to the
formula:
\begin{equation}
\eta^*\simeq\eta_c\frac{\ln(r_s/a)}{\ln(A_0/2\pi a)},
~~~~(a\exp(\eta/\eta_c) > r_s > A_0/2\pi),
\label{eq:eta}
\end{equation}
where $A_0 =4\pi\eta_cR^2/Q$ is the distance
between DNA turns on the surface of a sphere
if $\eta^*=\eta_c$
and $a$ is the radius of a DNA helix (see Fig.  \ref{fig:bead}).
One can show that $\eta^*=\eta_c$ if $r_s < A_0/2\pi$. 

\begin{figure}[htbp]
    \epsfxsize=5.5cm \centerline{\epsfbox{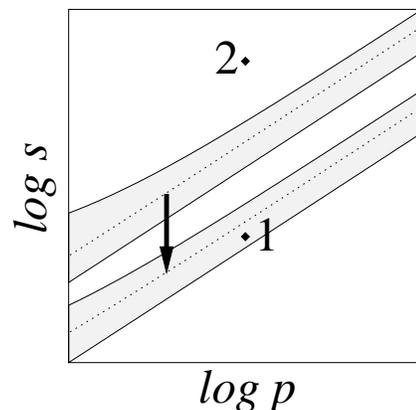}}
  \caption{The shift of isoelectric line and the ``neck"
  of phase diagram (gray)
  to lower $s$ as a result of the addition of monovalent salt. The points
  1 and 2 are used in Sec. IV.}
  \label{fig:twophase}
\end{figure}

It is the net charge density $\eta^*$ of DNA which
defines a renormalized isoelectric concentration
$s_{i}^*(p)= (\eta^*/\eta)s_i(p)$.  Thus, when $r_s$ decreases, the
phase diagram moves down together with $s_{i}^*(p)$ as shown
in Fig.  \ref{fig:twophase}. As we mentioned above, screening
also leads to some change of the width of condensation domain,
but this effect is negligible compared to the shift of the
isoelectric line.
The phase diagram presented above and its phenomenological
theory actually have a much broader applications
beyond the complexation of DNA with large spheres.
It was actually discovered in experiments of complexation 
of
oppositely charged proteins more than half century
ago\cite{deJong} but have not get a theoretical explanation.
Other applications are discussed in Sec. IV.
Here, we would like to mention only the well known phenomenon of
reentrant condensation of double-helix DNA
in solutions with small size multivalent counterions
with charge $Z \geq 3$ ($Z$-ions), such as spermine, a short
positive PE with $Z= 4$. At some critical concentration of
$Z$-ions, $s_c$, DNA abruptly condenses into large 
bundles\cite{Bloomfield}. (We use for $Z$-ions
the same notation as for spheres to emphasize complete
analogy at the phenomenological level).
Recently, it was discovered that at
another much larger critical $Z$-ion concentration, $s_d$, DNA bundles
dissolve again into free DNA 
helices\cite{Thomas,Livolant,Pelta,Raspaud,Raspaudnucl}. 
For spermine at very small
DNA concentrations, $s_c = s_c(p\rightarrow 0) = 0.025$~mM
and $s_d = s_d(p\rightarrow 0) = 150$~mM, 
if the monovalent salt concentration is not large.

A theory of reentrant condensation of DNA by spermine
was given in Ref. \onlinecite{NRS}. 
It is also based upon
two phenomenological parameters $\varepsilon$ and $s_0$, which have
meaning of the binding energy of DNA per $Z$-ion in the
aggregate
and the concentration of $Z$-ions which is in equilibrium
with neutralized-by-$Z$-ions DNA. By comparison
of $s_c$ and $s_d$ with theory,
they were found to be $\varepsilon \simeq 0.3k_BT$ and
$s_0 = 3.2$ mM. Ref.~\onlinecite{NRS}, however, dealt only with very small
$p$. The phenomenological theory of this
paper extends the theory of the spermine induced reentrant condensation
to the whole plane $(s, p)$.

Although DNA-spheres complexes and DNA with spermine have the
same phase diagram, there are important quantitative
differences between them. This is because,
as we will see in the next section,
the concentration $s_0$ decreases exponentially
with the binding energy of a $Z$-ion or a sphere with DNA.
For large strongly charged spheres, the binding energy of a sphere with
wrapping DNA can be so large that $s_0$ is extremely small.
Therefore, for the case of ``artificial chromatin", one most likely
deals with the ``neck" of the phase diagram
of the Fig. \ref{fig:phase} or,
in linear scale, with Fig. \ref{fig:lphase}a. 
This agrees with the phase diagram obtained
in Ref. \onlinecite{Keren} for DNA-large colloids complexation.
On the other hand, for  small $Z$-ions, the
energy of their binding to DNA is much smaller and $s_0$ is an easily
observable concentration.  In this case, one should see 
Fig. \ref{fig:lphase}b or the left side
(``the body") of the phase diagram of
Fig. \ref{fig:phase} instead. 
Indeed, the experimental diagram obtained for the
spermine~\cite{Raspaud} looks like the left half of Fig. \ref{fig:phase}.

The paper is organized as follows.
In Sec. II, we present a phenomenological theory of
the complexation of DNA with spheres.
Analytical formulae for
the critical concentrations $s_c(p)$ and $s_d(p)$ are derived and
details of the shape of the condensation and
decondensation transitions near $s_c(p)$ and $s_d(p)$ is discussed.
In Sec. III we present microscopic theory for parameters
$\varepsilon$ and $s_0$, for artificial chromatin
and study the role of screening by monovalent salt.
Onsager-Manning condensation on DNA wrapping around
a sphere is considered in Sec. IV.
Only in Sec. IV, we return to the discussion of other examples
of application of the same phase diagram.
In Sec. V, we discuss limits of applicability of our theory
of reentrant condensation. In the conclusion, we summarize
our results and discuss the possible implication of this
theory to the natural chromatin.

\section{Phenomenological theory and phase diagram of reentrant
condensation}

To begin with, let us first study a complex of a single DNA with
$N$ spheres. Its free energy can be written as
\begin{equation}
\label{eq:Fone}
f(N)=Q^{*\ 2}/2C + NE(N)~~.
\end{equation}
where
$Q^*=qN-Q=Q(N/N_i-1)$ is the net charge of
the complex ($q$ and $Q$ are the charges of one sphere
and a DNA molecule and $N_i=Q/q$ is the number of spheres necessary to 
neutralize a DNA helix).
Except for a specific value of the
sphere concentration $s$, the DNA-spheres complex is always 
charged and, therefore, has an extended shape to minimize its Coulomb
self-energy.  At a length scale larger than the average distance $l$ between
the spheres along the complex, the complex can be considered as a charged
cylinder with length $L$, radius $l$ (see Fig. \ref{fig:bead})
and linear charged density $Q^*/L$.
Then the capacitance $C$ of the complex is:
\begin{equation}
C=\frac{DL}{2\ln(r_s/l+1)}~,
\label{eq:capacitance}
\end{equation}
where $D$ is the dielectric constant of water and
monovalent salt in solution is treated in 
the Debye-H\"{u}ckel approximation with screening length $r_s$.

The first term in Eq. (\ref{eq:Fone}) is the standard
self-energy of a complex with net charge $Q^*$ and capacitance $C$.
The second
term is the correlation energy, which accounts for the
discreteness of the spheres charge at 
the length scale smaller than $l$ 
(in the beads-on-a-string structure, it is essentially the interaction
of a sphere with the DNA coil wound around it).
The correlation energy per sphere $E(N)$ is negative and,
in general, it 
is to be calculated from microscopic
theory. 
In this section, it is assumed to be known
and is used as an input parameter of the theory. Below,
we show that the parameter $s_0$
mentioned in the introduction is directly related to $E(N)$.

Given a DNA concentration $p$, we want to calculate the value of the
sphere concentration $s_c(p)$ and $s_d(p)$ where condensation and
decondensation of DNA-spheres complexes happen.  
To do so, let us consider the system in a
transition state where aggregates coexist with free complexes.
Let $x$ be the fraction of DNA in the aggregates, the
concentration of free DNA-spheres complexes in solution is then
$(1-x)p$. Correspondingly, the concentration of free spheres in
solution is $s-xpN_i-(1-x)pN$, where $xpN_i$ is the concentration
 of spheres
consumed by the neutral macroscopic aggregates, and $(1-x)pN$ is the
concentration of spheres bound to the free DNA-spheres complexes in the
solution. The free energy per unit volume of the system can then be
written as:
\begin{eqnarray}
\label{eq:F}
&&F(N,x;s,p)= (1-x)p f(N)
+xpN_i[E(N_i)+\varepsilon]+ \nonumber \\
&&k_BT\left[s-xpN_i-(1-x)pN\right]
        \ln\frac{\left[s-xpN_i-(1-x)pN\right]v_0}{e}.
        \nonumber \\
\end{eqnarray}
The first term in Eq. (\ref{eq:F}) is the free energy density of the free
complexes with $f(N)$ given by Eq.  (\ref{eq:Fone}). The second term
is the free energy density of the aggregates, $N_i\varepsilon$ is the energy
gained per complex by forming the aggregates (compared to a free
neutral isolated DNA-spheres complex in solution). This
energy gain originates from the
correlation-induced short range attraction between complexes
mentioned in the introduction ($\varepsilon$ is 
negative, see Fig. \ref{fig:2spheres}). Like
$E(N)$, $\varepsilon$ is to be calculated from
microscopic correlation theory, but in this section, 
it is considered as another
input parameter of the theory.  The third term in Eq.  (\ref{eq:F}) is
the free energy density of the concentration $s-xpN_i-(1-x)pN$ 
of left over free spheres in
solution. This concentration is assumed to be small, 
so that the solution of free spheres is ideal with
the normalizing volume $v_0$. 
The translational entropy of DNA is neglected in zeroth order
approximation.
This is valid if the DNA helix is long enough and the
concentration $p$ of DNA is not too small.

For given average concentrations $p$ of DNA and $s$ of spheres,
the state of the system can be found by minimizing the free energy
(\ref{eq:F}) with respect to the aggregate fraction $x$ and the number
of spheres $N$ bound to a free DNA-spheres complex. This gives:
\begin{eqnarray}
\label{eq:muCI}
&&k_BT\ln[(s-xpN_i-(1-x)pN)v_0]= \mu_c(N)+ \frac{qQ^*}{C} \\
&&N_i\varepsilon= 
 \frac{\left[Q(N/N_i-1)\right]^2}{2C}+(N_i-N)\times
 	\nonumber \\
&&~~~~~~~\left(k_BT\ln[(s-xpN_i-(1-x)pN)v_0]-\mu_c(N_i)\right)~. 
\label{eq:muPE}
\end{eqnarray}

Eqs. (\ref{eq:muCI}) and (\ref{eq:muPE}) have very simple physical 
meanings. Eq.
(\ref{eq:muCI}) 
equates the chemical potential of free spheres in
solution (the left hand side) and the electro-chemical potential of spheres
bound to the DNA-spheres complex (the right hand side). Here
$Q^*/C$ is the average electrostatic potential at
the surface of the complex
and $\mu_c(N)=\partial[NE(N)]/\partial N < 0$ is the
contribution to the chemical potential due to sphere correlations in
the complex\cite{Shklov003}.

Eq. (\ref{eq:muPE}) is the equilibrium condition for neutral
complexes.  The left hand side of Eq. (\ref{eq:muPE}) is the binding
energy per complex in the aggregates and the right hand side is the
chemical potential of a free DNA-spheres complex in solution. The 
latter chemical
potential is the sum of the complexes' self-energy (neglecting its
translational entropy) plus the entropy of $(N_i-N)$ spheres released
into solution. (In the dissolved state, a neutral complex releases
$(N_i-N)$ spheres charging itself to the charge $Q(N/N_i-1)$.) 

Before progressing further, it should be mentioned that,
both lengths $L$ and $l$ are, generally speaking, functions of the
number of spheres in the complex, $N$. 
A detail calculation on the
dependence of $L$ and $l$ on $N$ is given in Ref. \onlinecite{Shklov003},
where the structure of complexes far from isoelectric point 
(deep in the plus
and minus region of the phase diagram of Fig. \ref{fig:phase})
was considered.
In this paper, however, we concentrate on the phase diagram of the reentrant
condensation. In the vicinity of the condensation region, 
the complexes are almost neutral. Therefore, in the above minimization
of the free energy,
$L$ and $l$ (and hence the capacitance $C$)
are considered as independent of $N$ and equal to
their value for a neutral complex.
Indeed, if we consider explicitly the dependence of $L$, $l$ 
and $C$ on $N$, on the right hand side of Eq. (\ref{eq:muCI})
we get an additional term of the order $Q^*/Q$
compared to the second term. It is shown below that
$Q^*/Q$ is proportional to the square root of
the ratio
\begin{equation}
|\varepsilon C/qQ| \ll 1~,
\label{eq:theratio}
\end{equation}
where $\varepsilon$ is the condensate binding energy per sphere
and $qQ/C$ is the interaction energy of a sphere with the whole
DNA molecule. Thus, the dependence of $L$, $l$ and $C$ on $N$ can 
be ignored. 
The inequality (\ref{eq:theratio}) is justified because, 
as shown in Fig. \ref{fig:2spheres}, $\varepsilon$
involves only interactions with a small part of DNA molecule
winding around a spheres.
A more quantitative justification of this assumption is given in
the next section where microscopic theory of DNA-spheres interactions
in the complexes and the condensate is given.

If we define a sphere concentration $s_0$ as
\begin{equation}
\label{eq:ns0}
s_0=\exp(-|\mu_c(N_i)|/k_BT)/v_0~,
\end{equation}
Eq. (\ref{eq:muCI}) can be rewritten as
\begin{equation}
\label{eq:muCI2}
k_BT\ln\frac{s-xpN_i-(1-x)pN}{s_0}= 
\left(\frac{N}{N_i}-1\right)\frac{qQ}{C}~~.
\end{equation}
From this equation, one can see that, when the concentration of
free spheres, $s-xpN_i-(1-x)pN$, is greater than the concentration
$s_0$, $N/N_i-1$ is positive indicating a charge inversion effect.
Because Coulomb interaction between spheres is much larger than
thermal energy, $|\mu_c(N)|\gg k_BT$, $s_0$ is an exponentially small
concentration. The range of $s$ where DNA-spheres complexes are
overcharged, therefore, is easily accessible experimentally.  
One also sees that, without correlations, $\mu_c=0$, $s_0=1/v_0$ and charge
inversion can never be observed. This confirms that correlation is the
driving force for charge inversion.

Using Eq. (\ref{eq:muCI}) and the assumption that
the charge of a complex near condensation region is almost
zero so that
$\mu_c(N)\simeq\mu_c(N_i)$, we can rewrite Eq. (\ref{eq:muPE}) as
\begin{equation}
\label{eq:muPE2}
|\varepsilon|= \left(\frac{N}{N_i}-1\right)^2\frac{qQ}{2C}~.
\end{equation}

Equations (\ref{eq:muCI2}) and (\ref{eq:muPE2}) make 
up the core of our theory.
They contain only two phenomenological parameters $\varepsilon$ and
$s_0$. 
The former describes the strength of short
range attraction between complexes and the later describes the
strength of sphere-DNA binding, both depend on the
charge and size of spheres.
Knowing them, one can easily
calculate the fraction $x$ of DNA in the condensate
and the net charge $Q^*=Q(N/N_i-1)$ of free complexes
as functions of $s$ and $p$. Thus, the
whole phase diagram of the reentrant condensation can be constructed.

Let us consider the limit of small DNA concentration $p$. 
In this case, neglecting $p$ inside the logarithmic functions, 
we can rewrite Eqs. (\ref{eq:muCI2}) and (\ref{eq:muPE2}) as
\begin{eqnarray}
&&k_BT\ln\frac{s}{s_0}=\left(\frac{N}{N_i}-1\right)\frac{qQ}{C}~,\\
\label{eq:Rouzina}
&&\frac{|\varepsilon|}{k_BT}=\frac{k_BT C}{2qQ}\ln^2\frac{s}{s_0}~.
\end{eqnarray}
Obviously, there are two solutions $s_c$ and $s_d$ for 
Eq. (\ref{eq:Rouzina}):
\begin{equation}
\label{eq:p0}
s_{c,d}=s_0\exp\left(\mp\frac{1}{k_BT}\sqrt{|\varepsilon|
\frac{2qQ}{C}}\right) ~.
\end{equation}
The solution with the minus sign, $s_c < s_0$, corresponds to the
concentration of spheres at which DNA complexes start to condense 
forming aggregates. The other solution with the plus sign,
$s_d > s_0$, corresponds to the concentration of spheres at which the
aggregates dissolve again into free DNA-spheres complexes in solution.
In Fig. \ref{fig:phase} and Fig. \ref{fig:lphase}b, 
the concentrations $s_0$, $s_c$ and $s_d$ are
shown on the vertical axis. 

The meaning of the concentration $s_0$ is also transparent in this
$p\rightarrow 0$ limit. It corresponds to the sphere concentration at
which a free complex is neutral ($N=N_i$).

Let us now study the dependence of the threshold
condensation and decondensation
concentrations $s_c(p)$ and $s_d(p)$ on the DNA
concentration $p$.  Because the binding energy $\varepsilon$ 
is a constant, Eq.
(\ref{eq:muPE2}) shows that, in the regime of coexistence of
aggregates and free complexes, $N$ is a constant with respect to $p$
and $s$. Let us denote this constant by $N_c$ for the condensation
transition and by $N_d$ for the decondensation transition: 
\begin{equation}
\label{eq:N}
N_{c,d}=N_i\left(1\mp\sqrt{2|\varepsilon| C/qQ}\right)~.
\end{equation}
The deviations of $N_c$ and $N_d$ from $N_i$ is related to 
the effective charges of the DNA-spheres complexes when
they are in equilibrium with their aggregates:
\begin{equation}
Q^*_{c,d}=Q(N_{c,d}/N_i-1)=\mp Q\sqrt{2|\varepsilon|C/qQ}~.
\label{eq:Qcd}
\end{equation}

Together with the assumed inequality (\ref{eq:theratio}),
Eq. (\ref{eq:Qcd}) confirms that the net charge of a complex
near condensation region is very small compare
to the bare charge of the DNA.

Eq. (\ref{eq:muCI2}), in turn, leads to the conclusion that
in the coexistence regime, the concentration of free spheres
$[s-xpN_i-(1-x)pN_{c,d}]$ in solution is a constant independent of
$p$. These constants, of course, are equal to 
the concentrations $s_c$ and $s_d$
obtained above for the limit $p\rightarrow 0$.  Thus, for the
condensation transition,
\begin{equation}
\label{eq:cond}
s(p;x)-xpN_i-(1-x)pN_c=s_c~,
\end{equation}
and correspondingly, for the decondensation transition
\begin{equation}
\label{eq:decond}
s(p;x)-xpN_i-(1-x)pN_d=s_d~.
\end{equation}

Now one can easily find the threshold concentrations $s_c(p)$ (where
DNA-spheres complexes start to condense into bundles)
and $s_d(p)$ (where all the bundles
dissolve again). Putting $x=0$ in Eqs. (\ref{eq:cond})
and (\ref{eq:decond}) one gets:
\begin{eqnarray}
\label{eq:phasec}
s_c(p)=s_c+pN_c~,\\
\label{eq:phased}
s_d(p)=s_d+pN_d~.
\end{eqnarray}
Thus we have simple linear expressions for the condensation and
decondensation threshold sphere concentrations as functions of the DNA
concentration. These functions are plotted in Fig. \ref{fig:lphase}
by the solid lines.
The same phase diagram is plotted in log-log scale in Fig. \ref{fig:phase}
where the lines distinguishing the coexistence and the complete
condensation regions are omitted for clarity.

Putting $x=1$ in Eqs. (\ref{eq:cond}) and (\ref{eq:decond}), 
one can easily calculate the two sphere concentrations
between which the aggregates consume all the DNA-spheres complexes
(we are talking about very long DNA neglecting
its translational entropy):
\begin{eqnarray}
\label{eq:phasec1}
s^\prime_c(p)=s_c(p;x=1)=s_c+pN_i, \\
\label{eq:phased1}
s^\prime_d(p)=s_d(p;x=1)=s_d+pN_i. 
\end{eqnarray}
One concludes that the width of the range of concentration $s$ where
100\% DNA is aggregated, $s^\prime_d(p)-s^\prime_c(p)=s_d-s_c$, 
is a constant.
Remarkably, this region actually does not enclose the isoelectric
line $s_i=pN_i$. In Fig. \ref{fig:lphase}b, the complete condensation
region is shaded dark gray.

The sphere concentration $s_0(p)$ at which a free complex is neutral
can also be calculated in this model. Setting $N=N_i$ in Eq. 
(\ref{eq:muCI2}) gives
\begin{equation}
s_0(p)=s_0+pN_i~~.
\label{eq:s0p}
\end{equation}
As expected,
$s_0(p)$ lies in between $s^\prime_c(p)$ and $s^\prime_d(p)$ 
(compare  Eq. (\ref{eq:s0p}) with 
Eqs. (\ref{eq:phasec1}) and (\ref{eq:phased1})).
In the phase diagrams of Fig. \ref{fig:phase} and \ref{fig:lphase}, 
the concentration $s_0(p)$ is plotted by the dash line.
At large enough $p$, this concentration is close to the isoelectric point.
At small $p$, however, it saturates at the finite concentration $s_0$.

It is useful to study the
relative width of the condensation region:
\begin{equation}
\frac{\Delta s(p)}{s_0(p)}=
	\frac{s_d(p)-s_c(p)}{s_0(p)}=
	\frac{s_d-s_c+p(N_d-N_c)}{s_0+pN_i}~.
\end{equation}
At small $p$
\begin{equation}
\frac{\Delta s(p)}{s_0(p)} = \frac{s_d-s_c}{s_0}=
    2\sinh\left(\frac{1}{k_BT}\sqrt{
	|\varepsilon|\frac{2qQ}{C}}\right)~.
	\end{equation}
For large spheres, Coulomb interactions are much larger than
the thermal energy, the argument of $\sinh$ function is large,
so that this relative width is exponentially large. It is
shown as the ``body" of the phase diagram (Fig. \ref{fig:phase}).
On the other hand at large $p$
\begin{equation}
\frac{\Delta s(p)}{s_0(p)} = \frac{N_d-N_c}{N_i}=
	\sqrt{\frac{8|\varepsilon| C}{qQ}} \ll 1~.
	\label{eq:width}
\end{equation}

The narrowing of relative width $\Delta s(p)/s_0(p)$ with
growing $p$ from the ``body" to the ``neck" of the 
phase diagram is 
clearly seen in the log-log scale of the phase diagram 
(Fig. \ref{fig:phase}). 
(Note that the ``neck" encloses the isoelectric line
as Eqs. (\ref{eq:phasec}) and (\ref{eq:phased}) suggest.)
The width of the condensation region, of course,
always increases with $p$ as shown in Fig. \ref{fig:lphase}.

We would like to note that, because of the exponential dependence of the
concentration $s_0$ on the correlation chemical potential of spheres
$\mu_c$ (see Eq. (\ref{eq:ns0})), 
the accessible range of the diagram is very sensitive to the size
and charge of spheres.
When spheres are large and highly
charged, $|\mu_c|/k_BT$ is very large, $s_0$ is unrealistically small
and one can access only the right side of this phase diagram (the
``neck''). For spheres
of smaller size $s_0$ may be a
more reasonable concentration and one can
see also the left half of the phase diagram (``the body"). 

\begin{figure}
    \epsfxsize=9cm \centerline{\epsfbox{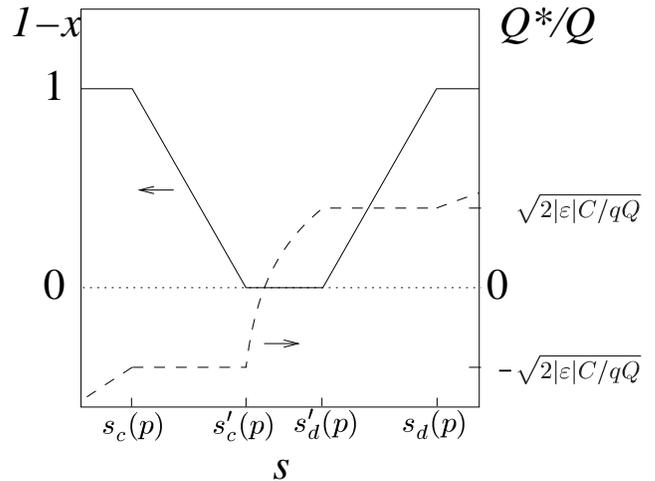}}
%
\caption{The fraction $(1-x)$ of free DNA in solution and the 
charge $Q^*$ (in units of $Q$) of a free complex as a function of the
concentration $s$ of spheres.}
\label{fig:fraction}
\end{figure}

Given the sphere and DNA concentrations, $s$ and $p$, 
one can also calculate the fraction $(1-x)$ of DNA dissolved 
in the solution.
\begin{equation}
1-x(s)=\left\{
	\begin{array}{ll}
		1& s<s_c(p), \\
		\parbox{1in}{$$\frac{-s+s_c+s_i}{s_i(1-N_c/N_i)}$$}& s^\prime_d(p) < s < s_d(p), \\
		0& s^\prime_c(p)<s<s^\prime_d(p), \\
		\parbox{1in}{$$\frac{s-s_d-s_i}{s_i(N_d/N_i-1)}$$}& s^\prime_d(p) < s < s_d(p), \\
		1& s>s_d(p).
	\end{array}
\right.
\end{equation}
This result is plotted in Fig. \ref{fig:fraction} by the solid line. 
The slopes
of $x(s)$ in the condensation coexistence region and in 
the decondensation coexistence region are equal in magnitude. They are
inversely proportional to the DNA concentration $p$. Therefore, at
small $p$, the transitions from $x=0$ to $x=1$
and vice versa are very abrupt. At large $p$, the coexistence region
is wider. This fact is shown in the phase diagram of Fig. \ref{fig:lphase}.
The widths of the light gray regions corresponding to coexistence 
are zero at $p \rightarrow 0$ and increase linearly with $p$. In
the limit of large $p$, the function $1-x(s)$ acquires a V-shape
form between $s_c(p)$ and $s_d(p)$.

Let us now consider how the charge $Q^*$ of a free complexes varies from
 negative at $s < s_0(p)$ to positive at $s > s_0(p)$.
When $s < s_c(p)$, Eq. (\ref{eq:muCI2}) shows that $Q^*$
grows logarithmically with increasing
$s$.  However, in the
first coexistent region, $ s_c(p) < s < s_c^\prime(p)$, 
the net charge of the complex $Q^*$
equal $Q_c^* < 0 $ and stays constant. 
In the second coexistence region, $s_d^\prime(p) < s < s_d(p)$,
the net charge is once more constant and but equal $Q_d^* =- Q_c^* > 0$.
 When $s > s_d(p)$, 
this net charge increases again logarithmically with $s$. In
Fig. \ref{fig:fraction}, this net charge is plotted by the dashed
line. In the region $s_c^\prime(p) < s < s_d^\prime(p)$, because
the condensate consumes all the free complexes, we have to go 
beyond Eqs. (\ref{eq:muCI2}) and (\ref{eq:muPE2}) to find $Q^*$.
Namely, in this interval we should take into account
translational entropy of free complexes.
This gives a new equation which replaces Eq. (\ref{eq:muPE2}):
\begin{equation}
N_i|\varepsilon| = \left(\frac{N}{N_i}-1\right)^2\frac{Q^2}{2C}-
	k_BT \ln[(1-x)pV_0],
\label{eq:muPE3}
\end{equation}
where $(1-x)p$ is the concentration of free complexes in equilibrium
with the condensate and $V_0$ is the normalizing volume for
the complexes.

Since we are interested in the variation of the net charge $Q^* = 
Q(N/N_i-1)$ with respect to $s$, let us eliminate $x$ from the two Eqs. 
(\ref{eq:muCI2}) and (\ref{eq:muPE3}). This gives:
\begin{eqnarray}
&&s - pN_i = s_0 \exp\frac{qQ^*}{k_BT C} + \frac{N_iv_0}{V_0}
s_0\left(\frac{N}{N_i}-1\right)\times
\nonumber \\
&&~~~~~~~~~~~~~~~~~~~~
	\exp\left(\frac{|\mu_c|-N_i|\varepsilon|}{k_BT}\right)
	\exp\frac{Q^{*\,2}}{2Ck_BT}.
\end{eqnarray}

For large $N_i$ (long DNA),
 the second term in the above equation is
exponentially small compared to the first term, thus one has:
\begin{equation}
Q^* = \frac{k_BTC}{q}\ln\left(1+\frac{s-s_0(p)}{s_0}\right)~.
\end{equation}
Using Eqs. (\ref{eq:p0}) and (\ref{eq:Qcd})
one can verify that $Q^*$ matches $Q^*_c$  and $Q^*_d$
at $s= s_c(p)$ and $s= s_d(p)$ respectively.

It is interesting to note that the region where the complexes
inverted charge is relatively narrow. Indeed, 
the slope at which $Q^*$ changes 
sign from negative to positive at $s_{0}(p)$ is
\begin{equation}
\frac{\partial Q^*}{\partial s}=\frac{Q}{s_0}\frac{k_BT2C}{qQ}~.
\end{equation}
Thus the characteristic width $\delta s = 
2Q^*_d/(\partial Q^*/\partial s ) $ over which 
charge inversion happens is exponentially smaller 
than $s_d - s_c$.

\section{Microscopic theory}

In the previous sections, we presented a description of the
correlation-driven reentrant condensation and
obtained an universal phase diagram based on two phenomenological
parameters $\varepsilon$ and $s_0$. These two parameters depend
on microscopic properties of the specific system considered.

In this section we use a microscopic theory 
of the beads-on-string structure (Fig. \ref{fig:bead})
for calculation of
$\varepsilon$, $s_0$ and their dependence on monovalent salt 
concentration. We consider large spheres
and, therefore, effectively work in the ``neck" of the
phase diagram (large $p$ case). 

First, for simplicity, we 
start from the theory for polyelectrolyte with linear charge density
$\eta \leq \eta_c=k_BT/e$.
The next section
considers highly charge polyelectrolyte
and the Onsager-Manning condensation which
renormalizes the net charge of DNA resulting in a shift of the
isoelectric line.

Because the DNA coil winding around a sphere almost
neutralizes the sphere charge,
the correlation chemical potential $\mu_c$ is essentially
the self energy of a bare free sphere
in solution which is almost totally eliminated 
in the complex\cite{Shklov003}. 
Thus one has
\begin{equation}
\mu_c = \left\{
		\begin{array}{ll}
	        -q^2/2DR& r_s > R, \\
			-q^2r_s/2DR^2& r_s < R.
	    \end{array}
		\right.
		\label{eq:muc}
\end{equation}
The parameter $s_0$ can be easily found with the help
of Eqs. (\ref{eq:ns0}) and (\ref{eq:muc}). One sees that,
at large $r_s$, the concentration $s_0$ is almost constant
but when $r_s$ decreases below $R$, $s_0$ increases exponentially
with $r_s$.

To calculate the correlation attraction energy $\varepsilon$, let
us use Fig. \ref{fig:2spheres}.
When two spheres touch each other, the density of the solenoid
at the touching region doubles. This leads to a gain in the
correlation energy of DNA turns (these correlations
develop at distances of the order of $A$ and should not be confused
with the correlation between spheres which determines $\mu_c$
and develops at distance of the order of $R$).

The correlation energy per unit length of the DNA can be estimated
as the interaction energy of the DNA segment with its stripe of background
(sphere) positive charge. Thus it is $-\eta^2\ln(R/A)/D$ for $r_s>R$
and $-\eta^2\ln(r_s/A)/D$ for $R>r_s > A$.
Correspondingly, when the density doubles ($A$ halves), the gain in 
the correlation
energy per unit length is $-\eta^2/D$ for $r_s > A$.
Similar effect takes place at $r_s <A$. In this case, 
each DNA interacts with the stripe of the width of $r_s$ of
the sphere surface  with energy $-q\eta r_s/2DR^2$.
When two spheres touch each other, it interacts with the stripe
on the other sphere as well, doubling the correlation energy.
Thus in this case, the correlation energy gain is $-q\eta r_s/2DR^2$.
Simple geometrical calculation shows that the total
length of DNA in the touching region (the region surrounded by
broken line in Fig. \ref{fig:2spheres}) is $R$. Therefore,
\begin{equation}
\varepsilon = \left\{
        \begin{array}{ll}
            -R\eta^2/D& r_s > A, \\
            -q\eta r_s/2DR& r_s < A.
        \end{array}
        \right.
		\label{eq:varepsilon}
\end{equation}

Using Eqs. (\ref{eq:muc}) and (\ref{eq:varepsilon}), we are
now in a position to discuss in detail the change of the
phase diagram with varying monovalent salt
concentration (varying $r_s$).
Let us consider
the following three regimes: $r_s > R$, $R > r_s >A$ and $r_s < A$

In the first regime, $r_s >R$, the parameters $s_0$ and $\varepsilon$
remain constant. The capacitance $C$ of a complex has the form:
\begin{equation}
C=DL/2\ln(r_s/R)~,~~~r_s > R~,
\label{eq:Clarge}
\end{equation}
where in Eq. (\ref{eq:capacitance})
the average distance $l$ between spheres has been replaced by $2R$,
because, near the condensation region, complexes are almost neutral
and spheres almost touch each other.
Therefore, Eq. (\ref{eq:Clarge}) and
(\ref{eq:muPE2}) show that, in the coexistence regions, the complex charge
$Q(N_{c,d}/N_i-1)$ increases logarithmically
with decreasing $r_s$.   Eq. (\ref{eq:width}) then shows that
the relative width of the condensation
region increases slowly in this regime:
\begin{equation}
\frac{\Delta s(p)}{s_0(p)}\sim\frac{2R\eta}{q}\sqrt{\frac{1}{\ln(r_s/R)}}
\ll 1 ~.
\label{eq:width1}
\end{equation}
Here, Eq. (\ref{eq:capacitance}) and the equality $Q/L=q/R$ were used.
Recall that we assume the number of turns, $m$, of DNA around a sphere
is large. Therefore $R\eta/q \simeq m^{-1} \ll 1$.
Decreasing $r_s$, however, leads to the contraction
of the condensation region
in the limit $p \rightarrow 0$. Indeed, Eq. (\ref{eq:p0}) shows that at
constant $\varepsilon$, the concentration $s_c$ increases 
and the concentration $s_d$ decreases.
Thus, overall, the condensation region decreases linearly
with decreasing $r_s$ at small $p$ and increases logarithmically 
at large $p$.

In the second regime, $R >r_s>A$, 
The concentration $s_0$ increases as $\exp(-r_s/R)$
while $\varepsilon$ remains constant.
The phase diagram moves upward with $s_0$.
Since $r_s < R$, the capacitance $C$ is
\begin{equation}
C=DLR/2r_s~,~~~r_s\ll R~.
\label{eq:Csmall}
\end{equation}
Thus, the relative width of the condensation region increases faster
than Eq. (\ref{eq:width1}):
\begin{equation}
\frac{\Delta s(p)}{s_0(p)}\sim\frac{2R\eta}{q}\sqrt{\frac{R}{r_s}}~.
\label{eq:width2}
\end{equation}
(Notice that Eq. (\ref{eq:width2}) matches Eq. (\ref{eq:width1})
at $r_s \sim R$).
At the same time, at small $p$, 
the condensation region shrinks faster $s_{c,d} \propto 
\exp(\mp r_s/R)$. Thus
in this regime, the change in shape the phase 
diagram continues the trend
in the first regime but with a faster pace. 

In the third regime, $r_s < A$, the energy $\varepsilon$ also starts 
to decrease linearly
with $r_s$. Eqs. (\ref{eq:muPE2}) and
(\ref{eq:p0}) therefore lead to the saturation in the 
change in $s_{c,d}$ and
$N_{c,d}$. The relative
width of the condensation region in the ``neck" saturates at:
\begin{equation}
\frac{\Delta s(p)}{s_0(p)}\sim\sqrt{\frac{2R\eta}{q}}~.
\label{eq:width3}
\end{equation}
(Notice that, because $A\sim R^2\eta/q$, Eqs. (\ref{eq:width2}) and
(\ref{eq:width3}) match each other at $r_s \sim A$.) On the
other hand, the concentration $s_0$ continues to increase 
exponentially with decreasing
$r_s$. One can say that in this region, the main effect of increasing
salt concentration is the growth of the ``body"
of the phase diagram Fig. \ref{fig:phase} at the expense of the ``neck".

\begin{figure}
\epsfxsize=8.0cm \centerline{\epsfbox{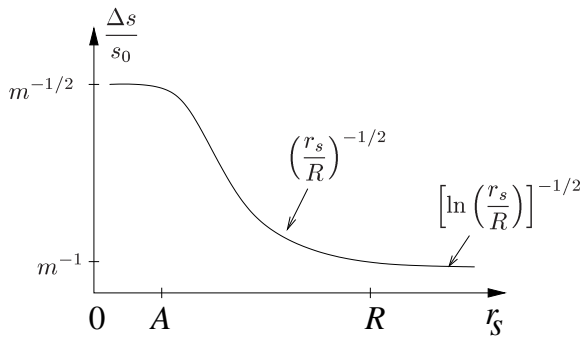}}
\caption{The relative width of the condensation region
of DNA-spheres complexes
as a function of the screening length. $m=q/\eta R \gg 1$ is the number
of turns of DNA neutralizing a sphere.}
\label{fig:width}
\end{figure}

The evolution of the width of the ``neck" as function of
$r_s$ is shown on Fig. \ref{fig:width}.

Concluding this section, we would like to justify 
quantitatively the assumption of Eq. (\ref{eq:theratio}),
which leads to conclusion that the charge of 
complexes is small near condensation region.
Using Eq. (\ref{eq:varepsilon}) for $\varepsilon$ and
Eqs. (\ref{eq:Clarge}) and Eq. (\ref{eq:Csmall}) for the complex capacitance
$C$ one gets:
\begin{equation}
\frac{|\varepsilon|C}{qQ} = \left\{
        \begin{array}{ll}
            (R\eta/q)^2/\ln(r_s/R) & r_s > R, \\
            (R\eta/q)^2 R/r_s&  R > r_s > A \simeq R^2\eta/q, \\
			(R\eta/q) & A >r_s.
        \end{array}
	\right.
	\label{eq:theratio2}
\end{equation}
One sees that in all cases, the ratio $|\varepsilon C/qQ|$
is at most $R\eta/q=m^{-1}$ where $m$ is the number of turns of
DNA coil on a sphere which we assume to be large.
Thus the inequality (\ref{eq:theratio}) is justified.

\section{Onsager-Manning condensation}
The theory presented so far is applicable to the 
complexation of polyelectrolyte
with linear charge density which does not exceed 
the so called Onsager-Manning critical charge
density $\eta_c=Dk_BT/e$. DNA helix is, however, a
highly charged polyelectrolyte with the bare negative
charge density $-\eta \simeq
-4.2\eta_c$. Condensation of positive monovalent ions on DNA,
therefore, must be taken into account. This
section deals with the problem of counterion condensation.

For a free DNA helix, the Onsager-Manning condensation
results in the net charge $-\eta_c$. However,
the net linear charge density of DNA winding around 
a charged sphere, $- \eta^*$, may differ from $-\eta_c$ 
due to the repulsion of counterions from the 
positive charge of the sphere. This phenomenon is called
counterion release\cite{Park}.
We show below that, for DNA coil which neutralizes 
a sphere, $\eta^*$ is given by Eq. (\ref{eq:eta}), so that
$ \eta \geq \eta^* \geq \eta_c$.
Similar to $L$ and $l$, the net charge density of DNA, $\eta^*$, can be
considered as independent on $s$, $p$ and $N$,
if one is interested only in 
the reentrant condensation domain of the phase diagram where the
charge of complexes are small.
Therefore, the theory used in previous section remains 
valid provided one replaces  $Q$ by $Q^*=Q\eta^*/\eta$, 
$N_i$ by $Q^*/q$ and  $s_i$ by $pQ^*/q$.

Remarkably, because the relative width of the condensation
domain studied in Sec. III is
small with or without screening, when Onsager-Manning 
condensation is taken into account the main effect
of screening in the phase diagram
is the decrease of $Q$ and, correspondingly, the isoelectric
concentration $s_i$.
The whole right part of diagram of Fig. \ref{fig:phase}
shifts in the direction of lower $s$ together with 
the renormalized $s_i$, as it is shown in  Fig. \ref{fig:twophase}.
To understand the implications of such 
shift, let us consider a solution 1, 
which is represented by the point $(s, p)$ below the curve $s_{c}(p)$
at very large $r_s$, when $ \eta^*$ is equal to the bare DNA
charge density $\eta$. If we add more monovalent salt and
reduce $r_s$, both curves $s_{c}(p)$ and $s_{d}(p)$ move down 
following the move of the
isoelectric line. As the result, our point 1
can be found inside the domain where complexes aggregate.
Thus condensation of negative (underscreened)
complexes of DNA and spheres 
can be induced by addition of salt.

On the other hand, if we start from the solution represented by
the point 2,
which at large $r_s$ is above the decondensation line 
$s_{d}(p)$, the result is completely different. 
In this case, addition of salt can not condense DNA-spheres
complexes. The origin of this asymmetry is that we are considering
large enough spheres with surface charge 
density smaller than that of DNA helix,
so that only monovalent cations experience 
the Onsager-Manning condensation.

It should be noted here that all the discussion above about
the change in the isoelectric line is relevant for the
``neck" of the phase diagram only. 
The expansion of the ``body" toward higher $p$ and $s$
with decreasing $r_s$ at $r_s < R$ mentioned in the
previous section is not affected by the
Onsager-Manning condensation.

Let us now derive Eq. (\ref{eq:eta}) for the net
linear charge density of the DNA due to Onsager-Manning
condensation. 
Let us denote by $c_1$ the concentration of monovalent salt 
in the solution
and by $c_s$ the concentration of monovalent ions condensed on the DNA
surface. As we know, for a free DNA in solution, counterions
condense on DNA making its net charge $-\eta_c$. 
The balance of the chemical potential of counterions in
solution and those condensed on the DNA helices, then reads
\begin{equation}
  \label{eq:monoCI1}
  k_BT\ln\frac{c_s}{c_1}=\frac{2e\eta_c}{D}\ln\frac{r_s}{a},
\end{equation}
where the left hand side is the entropy lost and  the right hand side
is the energy gained when a counterion condenses on the DNA surface,
$a$ is the radius of the DNA double helix (not to be confused with 
the radius $l$
of the complex, which, for almost neutral complexes where spheres
almost touch each other, is roughly the radius 
of one sphere, $l\simeq R \gg a$).

In the DNA-sphere complex, the positive charge of the sphere 
reduces
the amount of the counterions condensed on the DNA making the net
linear charge of DNA equal $-\eta^*$, instead of $-\eta_c$. The equilibrium
condition for the monovalent counterions now reads
\begin{equation}
  \label{eq:monoCI2}
  k_BT\ln\frac{c_s}{c_1}=\frac{2e\eta^*}{D}\ln\frac{A/2\pi}{a}-e\psi(0).
\end{equation}
Here $A/2\pi=2R^2\eta^*/Q$ plays the role of 
the local screening length at the DNA surface
($A$ is the distance between
subsequent turns of DNA around a sphere (see Fig. \ref{fig:bead})).
If $A/2\pi \gg r_s$ then $A/2\pi$ should be replaced by $r_s$. 
The average
electrostatic potential at the surface of a DNA-sphere complex 
is
$\psi(0)=[2Q^*(N/N_i-1)/DL]\ln(r_s/R)$. (Here and below, 
bearing in mind that we are working in the 
vicinity of isoelectric point we replaced $l$
by $R$).

Combining Eqs. (\ref{eq:monoCI1}) and (\ref{eq:monoCI2}), one gets
\begin{equation}
  \label{eq:monoCI}
  \eta_c\ln\frac{r_s}{a}=\eta^*\ln\frac{A}{2\pi a}-\frac{Q^*}{L}
        \left(\frac{N}{N_i}-1\right)
        \ln\frac{r_s}{R}~~.
\end{equation}
Near the condensation domain,
the charge of a complex $(N/N_i-1)$ is very small and
the second term in the right hand side
in Eq. (\ref{eq:monoCI}) is negligible.
(Indeed, according to eq. (\ref{eq:muCI2}), $Q^*(N/N_i-1)/L
\sim k_BTD/q \ll k_BTD/e =\eta_c$, and because $R \gg a$,
the second term is much smaller than $\eta_c\ln(r_s/a)$).
Using $A/2\pi=2\eta^*R^2/Q$ and substituting $\eta^*$ by $\eta_c$ in
$\ln(A/2\pi a)$,
one arrives at $N$-independent net linear charge density for DNA
given by Eq. (\ref{eq:eta}). This net charge 
decreases with decreasing $r_s$
(increasing salt concentration). When $r_s$ reaches $A/2\pi$, 
the net charge density $\eta^*$
reaches $\eta_c$. When $r_s$ is even smaller, 
$A/2\pi$ is replaced by $r_s$ and
$\eta^*$ saturates at the value of $\eta_c$.

The above derivation of $\eta^*$ is similar to that of
Ref. \onlinecite{Shklov0011} where the problem of adsorption
of DNA (with its Onsager-Manning condensed counterions)
on a charged surface was considered. However, the obtained net charge 
(Eq. (\ref{eq:eta})) is different with that of Eq. (7) of
Ref. \onlinecite{Shklov0011}. This is because
Ref. \onlinecite{Shklov0011} considers the case when 
DNA strongly overcharges
the surface, while this paper concerns with almost neutral
DNA-spheres complexes.

In the final paragraph of this section, we would like
to discuss the condensation of negative monovalent ions
(coions of DNA)
on the spheres. This happens if the surface charge 
density of the sphere is large enough and the screening
radius is not very large\cite{Chaikin}. 
Near the condensation domain, however, DNA coil almost neutralizes a sphere
eliminating its Coulomb potential at distances larger than $A/2\pi$.
Therefore,
most of coions are released from the sphere. There is still
a small amount of them condensed in the middle between
two DNA turns. However, the total charge
of these coions is much smaller than the total charge of DNA counterions
condensed on the DNA because, for $A\gg a$, DNA has a much
larger bare
surface charge density. Thus these coions are not important
in our calculation of the
isoelectric point. This also
means that
our calculation of $\varepsilon$ is valid.
On the other hand, $\mu_c$ can be modified when sphere condensation
is taken into account because the self energy of the sphere in
this case is smaller than that given by Eq. (\ref{eq:muc}). 
However, $\mu_c$  affects only $s_0$. For large spheres
where Coulomb interaction is much larger 
than $k_BT$, we always are in the ``neck" of the phase diagram
and $s_0$ is irrelevant. Therefore, the effect
of condensation of DNA coions on spheres is small.

\section{Other application of phenomenological theory
of reentrant condensation}

In Sec. II, we presented a theory of the phase diagram for
complexation of a long DNA helix with oppositely charged spheres.
It shows that correlation induced charge inversion
leads to reentrant condensation.
The phase diagram is described
by two parameters $\varepsilon$ and $s_0$.
Beside these two parameters, however, the phenomenological
theory does not use any
specific information about the structural and chemical properties of
the DNA. Therefore, this
theory is generic and can be used to describe
a broad range of systems experiencing reentrant condensation.
In this section, we would like to discuss briefly a number of
other systems of large opposite
charges which can be described by the same phase diagram of Fig.
\ref{fig:phase}.

Spheres can be replaced by particles of any shape
which we for brevity call $Z$-ions.
There is no need for them to be rigid. For example, they can be
star micelles or just oppositely charged shorter polyelectrolytes (PE),
more flexible than DNA, so that DNA can be considered rigid.
Still they will complex with DNA, for example, winding around it,
repel each other on DNA and form a kind of necklace.
Phenomenological parameters $s_0$, $\varepsilon$ can still be introduced
(but, of course, microscopic calculation of these parameters is
different from that of Sec. III) and
the phase diagram should look similar to Fig. \ref{fig:phase}
or Fig. \ref{fig:lphase}.

Let us discuss the limit when the $Z$-ion size is smaller than double
helix diameter so that they form two dimensional correlated liquid
on the surface of DNA. A well studied example of
this limit is the reentrant
condensation of DNA in the presence of spermine already
mentioned in Introduction.
Microscopically, it is different from the
beads-on-string systems, because in this case
DNA helices can be considered as rigid cylinders.
Condensation of DNA, of course, requires attraction between like-charged
cylinders. This attraction is related to the fact that $Z$-ions,
at the surface of DNA cylinders, strongly repel each
other and form a two-dimensional strongly correlated liquid.
At the place of the contact where the correlated liquids
of two touching DNA merge, the two-dimensional concentration of $Z$-ions
doubles. This leads to an energy 
gain~\cite{Rouzina,Mashl,Levin,Shklo006,Netz1} because
the correlation energy of a strongly correlated
liquid per $Z$-ion is known to be negative and
increases in absolute value with increasing $Z$-ion concentration.
In other words, two DNA
helices with strongly correlated liquid of $Z$-ions on their 
surfaces experience a
correlation-induced short range attraction.

It was predicted~\cite{Perel99,Shklov99} that in a
very weak DNA solution, the net charge of a DNA helix inverts sign at
a critical concentration $s = s_0$.  It has the meaning of
concentration of $Z$-ions in solution which is in equilibrium with the
strongly correlated liquid of $Z$-ions at the surface
of DNA (concentration of ``saturated
vapour" above strongly correlated liquid ).
Physics of this inversion of charge is also related
to correlations: when a new
$Z$-ion approaches already neutralized DNA it forms an
image in strongly correlated liquid, which attracts it.

Combining correlation induced charge inversion and
short range attraction, Ref.~\onlinecite{NRS} offered
an explanation for the origin
of the reentrant transition at small concentration of DNA.
At $s < s_0$ charges of two helices are negative.
Their absolute values decrease with increasing $s$ until at $s = s_c$,
where Coulomb repulsion looses to the correlation attraction and DNA
condenses in bundles. At $s > s_0$ the net charge of DNA becomes
positive and grows with increasing $s$. At $s= s_d$, the Coulomb repulsion
wins over correlation attraction and DNA bundles dissolve.
For $p \rightarrow 0$  Eqs. (\ref{eq:Rouzina}), (\ref{eq:p0})
were derived already in Ref.~\onlinecite{NRS}.
What is done in phenomenological theory of Sec. II
 can be considered as extension of the
theory of Ref.~\onlinecite{NRS} to arbitrary $p$.

Applicability of the phase diagram of Fig. 
\ref{fig:phase} of course, is not limited to DNA.
For all the broad spectrum of $Z$-ions, long
DNA helices, in turn, can be replaced by
other long, strongly charged PE. Experiment on
complexation in mixture of micelles and oppositely
charged PE~\cite{Dubin} is a good example.
In this experiment the total charge of micelles
was varied by changing the concentration
of the cationic lipid in the solution.
At some critical point the electrophoretic mobility
of complexes changes sign. In agreement with the above theory
measurements of dynamic light scattering and turbidity
(coefficient of light scattering)
show that complexes condense in bundles
in the same vicinity of the point
where mobility crosses over between two almost constant
positive and negative values. 

An interesting new case is that of
a weakly charged (one charge per $1/f \gg 1$ monomers) flexible PE.
Let us consider its complexation with small strongly charged spheres.
In this case, when PE winds around a sphere, the entropy of PE chain
prevents its collapse to the surface of the sphere.
As a result PE builds around the sphere layers of blobs with changing size,
similar to the one discussed in Ref.~\onlinecite{Rubinstein}.
The effective radius of a small sphere can be renormalized to a
larger value. The rest of theory of complexation
is similar to Ref.~\onlinecite{Shklov003} and this
paper so that our phase diagram works in this case, as well.

Complexation of a long PE with much shorter oppositely charged
PE in a water solution is important class of such problems.
Let us assume that the long PE is flexible
while the shorter one is rigid and stronger charged.
Then the long one sequentially wraps around molecules of the short one
so that rods-on-a-string system resembling
Fig.  \ref{fig:bead} is formed. Then the only change to be
made in the theory
of Ref.~\onlinecite{Shklov003} and this paper is to replace self 
energy of a
charged sphere by self energy of a charged rod. The case when the short
polymer is flexible (but still stronger charged) 
on the first glance seems to be more complicated.
However, away from isoelectric point short PE is either
underscreened or overscreened by the wrapping long one, so that
it has the rode-like shape. Similarly the whole complex is
charged, stretched and always resembles Fig.  \ref{fig:bead}.
Finally, if both PE are weekly charged but even the short one
is long enough, so that its charge is large,
they still form a complex resembling Fig.  \ref{fig:bead} with some
renormalizations of geometrical parameters related to
a hierarchy of blobs.

A long DNA helices or long PE  molecules
can be also replaced by any large macroions, for example, a
large colloidal particle, which adsorbs smaller
$Z$-ions. With appropriate correction for the
expression of the capacitance $C$ of the complex which,
in this case, has spherical instead of cylindrical shape,
our theory is applicable as well. An interesting practical
example of such system is solution of positive
(latex) spheres with short DNA oligomers with, say, 8 or 16
bases, which are adsorbed at the surface of spheres\cite{Grant,Gotting}.
Modified DNA can be delivered by such colloids as a drug. Therefore
question of stability of
such solutions at a given concentration of DNA is very
important.

It is interesting that in this case
latex spheres play the role of long DNA helix in the theory
of our paper, while short DNA oligomers play the role of spheres
or $Z$-ions.
Indeed, short DNA oligomers form two-dimensional correlated liquid
on the surface of colloid and can overscreen it.
 Simultaneously this liquid
provides attraction of almost neutralized colloids leading
to reentrant condensation around isoelectric point.
Reentrant condensation and change of the sign of electrophoretic
mobility in these systems were carefully studied
experimentally~\cite{Grant,Gotting}
and the results seem to agree with our theory.
Electrophoretic mobility of free complexes changes sign
in the very narrow vicinity of isoelectric point where
solution is unstable and easily coagulates.

Condensation of colloids with multivalent $Z$-ions was recently
observed in Monte-Carlo experiments\cite{Lobaskin} although
the concentration of $Z$-ions are not large enough to observe
resolubilization.

Another example of application of our phase diagram to almost spherical
macroions is the solution of nucleosomes with spermine\cite{Raspaudnucl}.
In the nucleosome the positive histone octamer (charge $\sim 160e$) is
strongly overcharged by DNA with charge $-292e$. Adsorption of
spermine cations on the surface of nucleosomes leads to 
nucleosome attraction and aggregation. Phase
diagram of this system looks like the body part of phase diagram of Fig.
\ref{fig:phase}. Electrophoretic mobility was measured for aggregates of
nucleosomes. With increasing concentration of spermine it changes sign
from negative to positive roughly in the middle between
condensation and decondensation curves $s_c(p)$ and $s_c(p)$ (in
logarithmic scale).

\section{Limitations of reentrant condensation.}

In the previous section, 
we saw that the combination of reentrant condensation and charge inversion
is a very general phenomenon. Now we would like to understand whether
this phenomenon takes place for any strongly charged and strongly
interacting electrolyte. We start from a theoretical
example of a system, which does not show this phenomenon.
This will help us to formulate limitations for
reentrant condensation and charge inversion.

Let us consider complexation of
an absolutely rigid cylinder-like negative PE
with small and strongly charged $Z$-ions~\cite{Shklo006}.
Let us assume that linear charge density of PE is $\eta$,
radius of the cylinder is $a$ and it is larger than radius of
$Z$-ion. We assume also that PE is 
not very strongly charged so that $Z/a\eta \gg 1$.
In this case, isolated complexes of a cylinder with adsorbed
$Z$-ions resemble one-dimensional Wigner crystal.
Two such  one-dimensional Wigner crystals strongly attract each other.
The deep minimum of Coulomb energy is always
provided by the condensed state, where cylinders form densely packed
background on the top of which $Z$-ions
form a three-dimensional strongly correlated liquid
with average distance between
$Z$-ions much larger than $a$
(they actually sit in pores of the background).
The binding energy of such strongly correlated liquid is much larger than
for free one-dimensional complexes.
Therefore, at large concentrations of PE and $Z$-ions,
this system is always
in condensed state and there is no way to see
reentrant condensation.

This (not very realistic) example emphasizes 
nontrivial nature of reentrant condensation and
charge inversion in the systems of DNA with large spheres or with spermine
(and other similar systems discussed above).
The destruction of aggregates by Coulomb repulsion of complexes 
at small distance from neutrality line (reentrant condensation)
is possible only if
the condensate binding energy per $Z$-ion, $\varepsilon$,
is smaller than the Coulomb interaction of a $Z$-ion with macroion, $qQ/C$.
Inequality $\varepsilon/ (qQ/C) \ll 1$ was used in Sec. II 
to simplify calculations.
If it is violated, the net charge $Q^*_{c,d}$
is not small, the neck widens
and can occupy the whole plane at large $s$
and $p$ eliminating condensation and decondensation transitions.
Of course, the requirement that interaction between $Z$-ion and DNA is large
($|\mu_c \gg k_BT|$), so that charge inversion can happen,
is also necessary for an observation of reentrant condensation.

Small values of $\varepsilon$ are realized for
DNA with large spheres or with spermine because in both cases
charges which form complexes very strongly screen each other.
For DNA with large spheres charge of a sphere is well screened by coiling DNA
so that spheres stick together only at the place they touch each other
(Fig. \ref{fig:2spheres})
and the energy $\varepsilon$ is very small. In the case of DNA
with spermine, $Z$-ions so strongly screen DNA
that attraction is developed only in the small area where two
DNA touch and the energy $\varepsilon$ is relatively small.

\begin{figure}
    \epsfxsize=7cm \centerline{\epsfbox{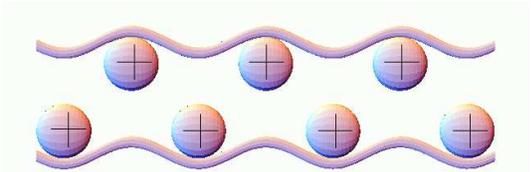}}
	\caption{Complexation of DNA and spheres with
	intermediate value of charge and size. Screening
	is not complete as in the case of large spheres
	or multivalent ions.}
	\label{fig:semiflexible}
\end{figure}

Between systems of DNA with large spheres
and DNA with spermine there are many others
with intermediate size and charge of spheres.
One can easily imagine an intermediate system,
where on one side spheres are not large and strongly charged enough
that DNA behaves as a flexible PE, and on the other
side they are strongly enough charged so that
they form something like one-dimensional Wigner crystal on its surface.
In such a case, mutual screening of DNA and spheres is not as complete as
in the case 
of DNA with large spheres or with spermine 
(see Fig. \ref{fig:semiflexible} or Ref. \onlinecite{Netz}). 
Therefore, inequality $|\varepsilon|/ (qQ/C) \ll 1$ can be violated and
reentrant condensation may not materialize.

Let us give an example of a different, well-studied system where 
this happens. Consider
a solution cationic lipid membranes and long double helix DNA.
Each membrane, in principle, may complex with many DNA rods which
lie on it equidistantly parallel to each other, forming a kind of
one-dimensional
Wigner crystal. (In such a complex DNA plays the role of $Z$-ion)
These complexes may be undercharged and overcharged.
They can also condense in three-dimensional aggregates, 
where layers of equidistantly positioned parallel DNA rods alternate
with membranes~\cite{Radler}.
Such aggregates have potential applications in gene therapy as non-viral
gene carriers.
In spite of liquid nature of membranes, which permits some
redistribution
of lipid charge inside membrane to screen DNA,
attraction between different complexes
in this crystal is so strong that situation for a reentrant
transition in this case is marginal ($|\varepsilon|C/qQ \sim 1$).
It seems from experimental data 
that indeed aggregates exist even far from the isoelectric
point~\cite{Radler}.

For the colloidal spheres covered by short DNA
mentioned in Sec. V, binding is related to the small area of contact
and energy $\varepsilon$ is small again.
Let us, instead of short DNA, consider small positive
colloidal particles.
If their size and charge are much smaller than that of larger 
negative ones
we still get reentrant condensation like with the short DNA.
But what happens when their charges and sizes
become comparable is not obvious. 
Therefore it is interesting to consider a simpler case of
a solution of positive and negative spherical
macroions with same absolute values of charges, $q$,
same radius, $R$ and with concentrations $c_+$ and $c_-$ respectively.
Let us talk about large enough $c_+$ and $c_-$,
when Coulomb energy of interaction of two touching spheres,
$E = q^2/2DR$, completely dominates the entropy.
It is obvious then that at $c_+ =  c_-$
such system tends to self assemble into NaCl-like crystal and
gain the Madelung energy  $-1.74E$ per negative sphere.
Away from the isoelectric point situation is less trivial.
Let us consider for simplicity a solution with
$c_+ = 2 c_-$ (neutrality is provided
by monovalent negative ions).
In this case one has to compare energies of
two states. The first one consist of
NaCl-like crystal consuming concentrations $c_-$ of
both positive and negative spheres.
The rest $c_-$ positive spheres are free.
Such state has the energy $-1.74Ec_-$
per unit volume of solution.
The second state consists only of $c_-$ triplets, in which two
positive spheres are attached on opposite sides to  a
negative one. Energy of such a complex is $-1.5E$.
We see that second state looses to the first.
It can be shown that at any
$c_+/c_-$, the largest possible fraction of
spheres always self assembles in neutral
NaCl-like crystal. Thus we are coming to conclusion that triplets
(which are positive
at $c_+ > c_-$ and negative at $c_+ <  c_-$) do not form.
Triplets can be considered
as analogs of free and charged DNA-spheres necklaces.
So in this case, neutral condensed state
always dominates and there is no reentrant condensation at all.

\section{Conclusion}

In this paper, 
we presented a phenomenological theory of reentrant condensation.
The theory is applicable not only to the solution of DNA and spheres
but also to a much broader range of systems.

Here, we would like to discuss possible
implications of our theory for the natural chromatin
assuming that the beads-on-a-string structure of 10 nm chromatin
fiber is indeed determined by electrostatic interactions.
In this case, the finite distance between nucleosomes
tells us that the whole 10 nm fiber is charged.
Electrophoretic experiments\cite{Fletcher} show that
in low salt conditions, the net charge of 10 nm fiber is negative.
(By the net charge we mean bare charges of DNA and histones plus
charges which are Onsager-Manning condensed on
them and, therefore, are bound to the fiber with the
energy larger than $k_BT$).
An additional argument for the negative sign of the charge is that
the increasing salt concentration condenses it
into the 30 nm fiber. This means
that 10 nm fiber obeys the scenario discussed in Sec. V
for the point 1 of  Fig.  \ref{fig:twophase}.

Looking at the phase diagram of Fig. \ref{fig:twophase}
we can ask whether one can increase concentrations of
both octamers and salt much further so that
that the point ($s, p$) becomes 
higher than the decondensation curve $s_d(p)$
and instead of 30 nm fiber or 
higher order structures we can
get a {\it positive} stable 10 nm fiber.
This is not easy because free octamers decay into
smaller histone dimers or tetramers. To keep equilibrium
concentration of free octamers $s$ in solution at the necessary
level, concentrations of histones
should be large. In spite of this difficulty we
suggest to try to create and study a positive 10 nm fiber.

\acknowledgements
We are grateful to A. Yu. Grosberg, V. Lobaskin, E. Raspaud,
I. Rouzina, U. Sivan and J. Widom for useful discussions. 
This work is supported by NSF DMR-9985785.

\end{multicols}
\end{document}